\documentclass{article}

\usepackage{arxiv}

\usepackage[utf8]{inputenc}
\usepackage[T1]{fontenc}
\usepackage{hyperref}
\usepackage{url}
\usepackage{booktabs}
\usepackage{amsfonts}
\usepackage{amsmath}
\usepackage{amssymb}
\usepackage{amsthm}
\usepackage{nicefrac}
\usepackage{microtype}
\usepackage{graphicx}
\usepackage{natbib}
\usepackage{doi}
\usepackage{algorithm}
\usepackage{algpseudocode}

\graphicspath{{Fig/}}

\hypersetup{
    colorlinks=true,
    citecolor=blue,
    linkcolor=blue,
    urlcolor=blue
}

\theoremstyle{plain}
\newtheorem{theorem}{Theorem}
\newtheorem{proposition}[theorem]{Proposition}

\theoremstyle{definition}

\title{BaySC: Uncovering Tissue Architecture in Spatial Multi-Omics\\via Probabilistic Spatial Clustering}

\author{
  Xin Li$^{1}$ \quad
  Xiaofei Dong$^{1}$ \quad
  Zhenke Duan$^{1}$ \quad
  Lulu Shang$^{3}$ \quad
  Xiao Wang$^{4}$ \quad
  Xinyuan Song$^{5}$ \\[4pt]
  \textbf{Hanwen Ning}$^{1,2,\dagger}$ \quad
  \textbf{Guanyu Hu}$^{6,\dagger}$ \\[8pt]
  $^{1}$School of Statistics and Mathematics, Zhongnan University of Economics and Law, Wuhan 430073, China \\
  $^{2}$Innovation and Talent Base for Digital Technology and Finance, Zhongnan University of Economics and Law \\
  $^{3}$Department of Biostatistics, The University of Texas MD Anderson Cancer Center, Texas 77030, USA \\
  $^{4}$Department of Statistics, Purdue University, Indiana 47907, USA \\
  $^{5}$Department of Statistics, The Chinese University of Hong Kong, Hong Kong, China \\
  $^{6}$Department of Statistics and Probability, Michigan State University, MI 48824, USA \\[6pt]
  \texttt{ninghanwen@gmail.com} \quad \texttt{huguanyu@msu.edu} \\
  $^\dagger$Corresponding authors
}

\renewcommand{\shorttitle}{BaySC: Probabilistic Spatial Clustering for Spatial Multi-Omics}

\hypersetup{
  pdftitle={BaySC: Uncovering Tissue Architecture in Spatial Multi-Omics via Probabilistic Spatial Clustering},
  pdfauthor={Xin Li, Xiaofei Dong, Zhenke Duan, Lulu Shang, Xiao Wang, Xinyuan Song, Hanwen Ning, Guanyu Hu},
  pdfkeywords={Bayesian Spatial Domain Detection, Markov Random Field, Mixture of Finite Mixtures, Spatial Omics},
}

\begin{document}
\maketitle

\begin{abstract}
\textbf{Motivation:} Spatial domain identification requires jointly modeling
molecular signatures and physical coordinates, yet current tools frequently
over-smooth biological boundaries, require user-specified cluster numbers,
and lack principled multimodal integration.

\textbf{Results:} We introduce BaySC, an integrative Bayesian spatial
clustering framework for spatial domain identification. BaySC inherently
learns the true number of spatial domains from the data by employing a
Mixture of Finite Mixtures (MFM) prior. Tissue topology is modeled via a
Markov Random Field (MRF) applied to discrete cellular assignments, a
strategy that enforces local spatial coherence without distorting the
underlying gene expression features. This enables BaySC to accurately map
contiguous tissue layers as well as geographically scattered,
transcriptionally identical cell populations. Furthermore, BaySC handles
spatial multi-omics data through a weighted log-likelihood fusion mechanism
executed via Gibbs sampling. This approach assigns interpretable weights to
each modality, allowing users to quantify the biological relevance of
different data layers to the final tissue map. Validated across ten
single-modal spatial transcriptomics and two spatial multi-omics datasets,
BaySC yields highly interpretable probabilistic outputs. It demonstrates
competitive accuracy on standard clustering metrics and consistently
outperforms existing tools in preserving spatial topography, as measured by
spatially-aware Adjusted Rand Index (spARI).

\textbf{Availability and implementation:} The code is available at
\url{https://github.com/lixin0304/BaySC}
\end{abstract}


\section{Introduction}

The emergence of spatial transcriptomics (ST) has enabled the systematic, transcriptome-scale measurement of gene expression while precisely preserving the in situ spatial microenvironment of tissues \citep{moses2022museum, rao2021exploring}. This technological advance has fundamentally expanded our ability to interrogate complex living systems across diverse biological contexts \citep{chen2023spatiotemporal,jorstad2023transcriptomic,hunter2021spatially}. Against this backdrop, spatial domain identification, which delineates functional tissue regions exhibiting both transcriptomic coherence and spatial continuity by jointly integrating molecular expression profiles with physical location information, has emerged as a central pillar of ST data analysis \citep{hu2021spagcn, dong2022deciphering, long2023graphst, singhal2024banksy}. The outputs of spatial domain identification directly support various downstream bioinformatic tasks \citep{gulati2024profiling, dries2021giotto, tian2024spavae}.

Despite substantial progress, spatial domain identification continues to face several profound analytical challenges. First, ST measurements are typically compromised by limited sequencing depth and stochastic dropout, rendering the resulting count matrices intrinsically high dimensional, sparse and noisy. Stable statistical modeling of such data requires highly effective representation and denoising strategies \citep{wang2022sprod, tian2024spavae}. Second, from the perspective of heterogeneous data integration, gene expression profiles are high-dimensional sparse vectors, whereas physical spatial coordinates are low-dimensional continuous variables. Coupling these intrinsically different data structures within a unified probabilistic framework remains a nontrivial statistical hurdle \citep{chen2025smopca, li2025spallm}. Third, driven by the rapid evolution of spatial multi-omics technologies, platforms such as MISAR-seq and Spatial-CITE-seq now permit the simultaneous capture of multiple molecular modalities within the same tissue section, including transcriptomics, chromatin accessibility (ATAC), and protein abundance (ADT) \citep{jiang2023simultaneous, liu2023spatialciteseq}. However, the relative contributions of these modalities to defining tissue architecture are often spatially heterogeneous. This shifting modality importance imposes substantial demands on model design, requiring adaptive integration mechanisms \citep{liu2025instinct, long2024deciphering}. Finally, the true number of spatial domains in real biological tissues is generally unknown and highly data-dependent. Pre-specifying the number of clusters, which is required by most current methods, remains biologically unjustified and frequently introduces systematic bias into downstream inferential analyses \citep{zhao2021spatial, miller2018mixture}.

Existing methods for spatial domain identification have largely evolved along a few representative technical trajectories, each carrying distinct methodological limitations. Graph neural network (GNN)-based approaches, exemplified by SpaGCN \citep{hu2021spagcn}, STAGATE \citep{dong2022deciphering}, GraphST \citep{long2023graphst}, and Giotto \citep{dries2021giotto}, typically construct a spatial adjacency graph and aggregate neighborhood expression information using graph convolution or graph attention, followed by standard clustering algorithms (e.g., K-means, mclust, or Louvain). BANKSY \citep{singhal2024banksy} concatenates intrinsic cellular expression features with neighborhood-averaged features to capture the local microenvironment \citep{li2025spallm, tian2024spavae}. A fundamental limitation of these approaches is that their spatial integration mechanism performs spatial smoothing directly on continuous expression features. Consequently, when the data are highly noisy or true biological boundaries are sharp, this feature-level smoothing inevitably leads to ``over-smoothing'', blurring transcriptionally distinct boundaries and producing spurious intermediate states that do not correspond to true biological structures. BayesSpace \citep{zhao2021spatial} offers a fully Bayesian framework that models spatial continuity via a Markov random field (MRF) prior on discrete cluster labels \citep{singhal2024banksy, miller2018mixture}. However, BayesSpace is structurally restricted to single-modality RNA data, which limits its performance.

For multimodal integration, SpatialGlue \citep{long2024deciphering}, SpaMI \citep{gao2025graph}, soFusion \citep{yu2025sofusion}, INSTINCT \citep{liu2025instinct}, and SMOPCA \citep{chen2025smopca} rely on deep attention mechanisms, contrastive learning, or representation fusion. While offering immense flexibility, these deep learning frameworks share several critical limitations: they treat the number of clusters as a rigid, user-provided input rather than an inferential target and lack principled mechanisms for quantifying cell-level clustering uncertainty \citep{tian2024spavae, dong2025spatranslator}. Achieving a principled equilibrium between statistical rigor, flexible multimodal integration, adaptive cluster-number inference, and controllable spatial regularization remains a highly pressing unmet need.

To resolve these limitations, we propose Bayesian Spatial Clustering (BaySC), a unified Bayesian spatial clustering framework tailored for spatial transcriptomics and multi-omics data. BaySC computes pairwise cosine similarities from low-dimensional embeddings of each modality \citep{moses2022museum, cusanovich2018atlas, liu2023spatialciteseq}, and employs the Fisher Z-transformation to map these similarities onto the real line. This design performs robust, modality-agnostic statistical standardization prior to inference, effectively normalizing depth-related scale differences \citep{wang2022sprod, chen2025smopca, long2024deciphering}. At its core, BaySC replaces the classical Chinese restaurant process with a Mixture of Finite Mixtures (MFM) prior \citep{miller2018mixture}, which leverages marginal likelihood ratio mechanisms to dynamically suppress redundant domains, enabling statistically consistent, data-driven inference of the true number of clusters. Furthermore, BaySC introduces an MRF constraint \citep{zhao2021spatial, singhal2024banksy} that strictly modulates the discrete label allocation probabilities during Gibbs sampling, fundamentally bypassing the feature-blurring artifacts of GNNs. Multimodal evidence is seamlessly integrated via a weighted linear summation of log-likelihoods \citep{chen2025smopca}, and the optimal spatial smoothing parameters are automatically calibrated using the modified deviance information criterion (mDIC) \citep{spiegelhalter2002bayesian}.

BaySC contributes significant advancements to both statistical methodology and biological practice:
\begin{itemize}
    \item \textbf{Statistical Innovations:} We establish a principled generative model that simultaneously infers the optimal number of spatial clusters via an MFM prior and calibrates spatial regularization via mDIC, removing arbitrary manual tuning. Furthermore, by placing the MRF constraint strictly on the allocation of discrete latent labels rather than on continuous observation features, BaySC encourages spatial contiguity among neighboring cells while fully preserving the capacity to cluster geographically disjoint but transcriptionally homogeneous structures without distorting the underlying data geometry.
    \item \textbf{Biological Insights:} The Bayesian formulation provides a level of interpretability unavailable in deep learning frameworks. The modality-specific likelihood weights offer a quantifiable metric to determine which molecular layer (e.g., epigenomic versus transcriptomic) derives the organization of specific tissue regions. Crucially, by extracting point estimates via the Dahl method \citep{dahl2006model, dahl2020salso}, BaySC yields exact posterior co-membership probabilities for every cell \citep{tian2024spavae}. This provides critical biological uncertainty quantification, enabling researchers to confidently identify transitional cellular subpopulations, map continuous developmental gradients, and dissect ambiguous domain boundaries within complex tissues.
\end{itemize}

Through comprehensive benchmarking across three single-modality datasets and two multimodal datasets, BaySC achieves clustering performance that is highly competitive with the strongest neural baselines on standard external metrics (ARI, AMI, NMI) \citep{gulati2024profiling}. Most importantly, BaySC consistently outperforms all competing approaches on the spatially aware metric spARI, definitively demonstrating its superiority in preserving spatial consistency and true domain boundaries. BaySC also establishes an extensible probabilistic foundation for future studies into temporally dynamic tissues and uncertainty-aware experimental designs.

\begin{figure*}[!t]
\centering
\includegraphics[width=0.9\linewidth]{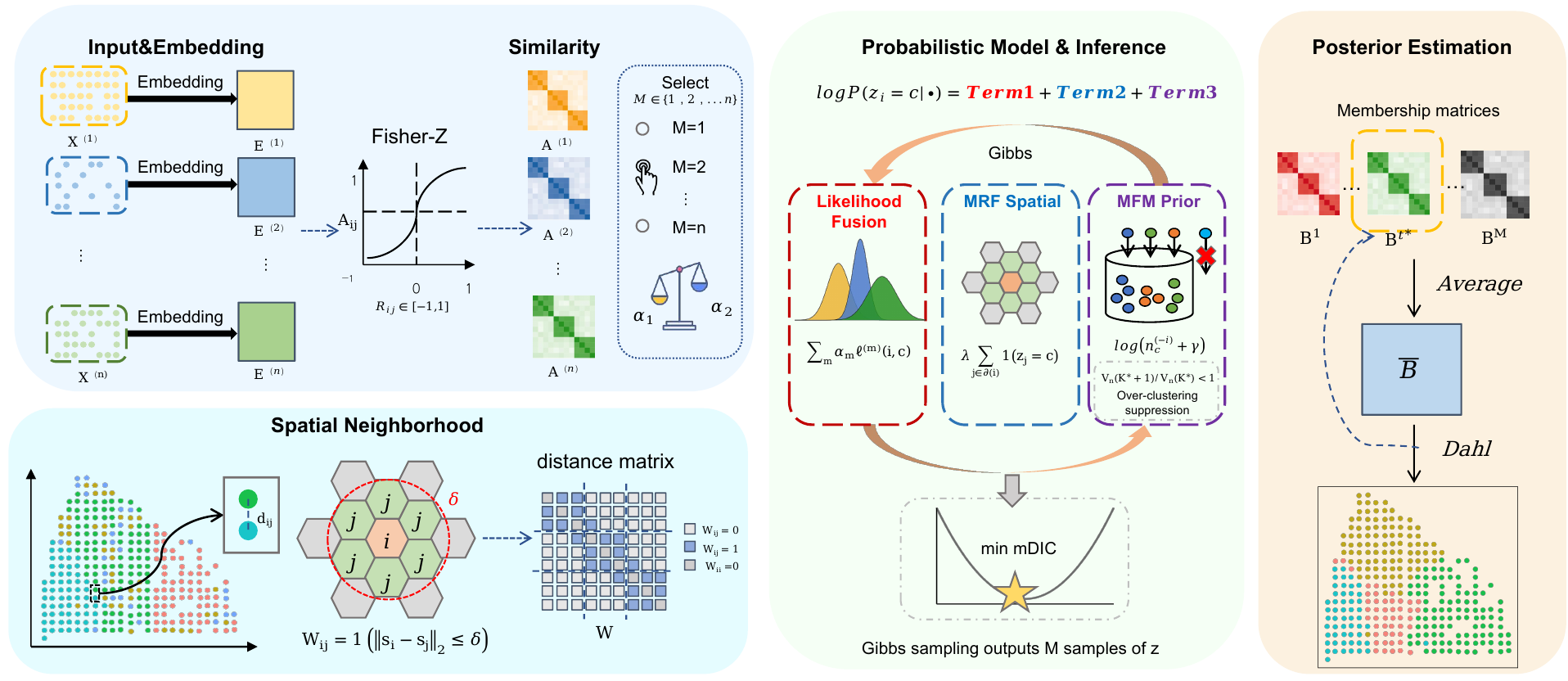}
\caption{Overall framework of BaySC.
\textbf{(Left)}~Raw count matrices from each molecular modality
are embedded into low-dimensional representations, from which
pairwise similarity matrices are derived via Fisher-Z
transformation. A binary spatial neighborhood matrix $\mathbf{W}$
is simultaneously constructed from physical spot coordinates.
\textbf{(Middle)}~A collapsed Gibbs sampler updates cluster
assignments by jointly balancing multimodal data likelihood,
MRF spatial coherence, and the MFM clustering prior. The spatial
smoothing parameter $\lambda$ is automatically calibrated by
minimizing mDIC.
\textbf{(Right)}~Posterior co-membership matrices collected
across Gibbs iterations are averaged and the representative
sample is selected via Dahl's method as the final domain
assignment.}
\label{fig:workflowBaySC}
\end{figure*}

\section{Materials and Methods}
\label{sec:methods}

\subsection{The Framework of BaySC}
\label{sec:overview}

We propose BaySC, a fully Bayesian integrative framework designed for robust spatial domain identification across spatial multi-omics datasets. By synergistically leveraging multiple molecular modalities, BaySC preserves tissue-level spatial architecture while automatically inferring the number of distinct spatial domains. The workflow encompasses four sequential stages, as illustrated in Figure~\ref{fig:workflowBaySC}.

\textbf{Step 1: Data Preprocessing and Graph Construction.}
Raw count matrices from each molecular modality are projected into low-dimensional latent embeddings, from which cell--cell similarity matrices are computed via cosine similarity and stabilized by Fisher-Z transformation. Physical tissue coordinates are independently used to construct a binary spatial neighborhood matrix. Together, these two matrix types constitute all inputs to the generative model.

\textbf{Step 2: Bayesian Hierarchical Modeling.}
BaySC formulates a Bayesian hierarchical model over the similarity matrices. The observation layer employs a Gaussian Stochastic Block Model (SBM) with Normal-Gamma conjugate priors to capture intra- and inter-domain molecular relationships. The clustering layer is governed by a joint prior combining an MFM---for data-driven inference of the domain number---and an MRF---for penalizing spatially isolated assignments to enforce structural contiguity.

\textbf{Step 3: Posterior Inference via Collapsed Gibbs Sampling.}
Inference is executed via a collapsed Gibbs sampler that analytically marginalizes out all continuous block parameters under Normal-Gamma conjugacy, restricting sampling to discrete domain labels. Multimodal evidence is fused through a weighted log-likelihood during label updates, while a specialized marginal likelihood governs new-domain proposals. The Dahl method resolves label switching and extracts a posterior point estimate, and the resulting co-membership matrix provides cell-level assignment uncertainty.

\textbf{Step 4: Data-Driven Hyperparameter Calibration.}
The spatial penalty ($\lambda$) and neighborhood threshold ($\delta$) are jointly selected via mDIC over a predefined grid. The mDIC applies a BIC-consistent complexity penalty to guard against over-clustering. The grid explicitly includes a $\lambda = 0$ baseline, allowing the model to deactivate spatial regularization when the data do not support it.

The complete BaySC procedure is also presented by Algorithm 1 in Supplementary Note 1.

\subsection{Data Representation and Similarity Construction}
\label{sec:step1}

Consider a tissue section containing $n$ cells with spatial coordinates $\mathbf{s}_i \in \mathbb{R}^2$, $i = 1, \ldots, n$. Each cell is associated with molecular measurements from one or more modalities, including RNA transcriptomics, ATAC chromatin accessibility, and protein abundance (ADT), in any combination. Step~1 transforms the heterogeneous raw multi-omics data into numerical inputs amenable to the probabilistic model. This process comprises three components: modality-specific embedding, cell--cell similarity matrix construction, and spatial neighborhood matrix construction.

\textbf{Modality-specific embedding.}
For each modality $m$, the raw count matrix (cells $\times$ features) is compressed through a standard preprocessing pipeline into a low-dimensional embedding matrix $\mathbf{E}^{(m)} \in \mathbb{R}^{n \times d_m}$, where $d_m$ is the retained embedding dimensionality. Specifically, the RNA modality undergoes library-size normalization, $\log(1+x)$ transformation, highly variable gene selection, and gene-wise $z$-score standardization, followed by PCA; the ATAC modality undergoes binarization and TF-IDF transformation, followed by truncated SVD with removal of the first component that predominantly captures sequencing depth; the protein/ADT modality undergoes centered log-ratio (CLR) transformation followed by PCA. Finally, cell-wise $z$-score standardization ($\mathrm{ddof}=1$) is applied uniformly across all modalities to eliminate differences in numerical scale.

\textbf{Cell--cell similarity matrix.}
Given the standardized embedding matrix $\mathbf{E}^{(m)}$, the cosine similarity between cells $i$ and $j$ is defined as:
\begin{equation}
    R^{(m)}_{ij}
    = \frac{1}{d_m}\,
      \bigl(\mathbf{E}^{(m)}\bigr)_i
      \cdot
      \bigl(\mathbf{E}^{(m)}\bigr)_j^{\!\top}.
\end{equation}
To map the bounded similarity values onto the entire real line with approximately homogeneous variance~\citep{hu2021spatially}, the Fisher-Z transformation is applied:
\begin{equation}
    A^{(m)}_{ij}
    = \operatorname{arctanh}\!\bigl(R^{(m)}_{ij}\bigr)
    = \frac{1}{2}\,\log\frac{1 + R^{(m)}_{ij}}{1 - R^{(m)}_{ij}}\,.
\end{equation}
Before transformation, $R^{(m)}_{ij}$ is clipped to $(-0.9999,\,0.9999)$ to prevent numerical overflow. The diagonal entries $A^{(m)}_{ii}$ are retained for use in the new-domain proposal (see details of Step~3).

\textbf{Spatial neighborhood matrix.}
Based on the coordinates $\mathbf{s}_i$, a binary spatial neighborhood matrix $\mathbf{W} \in \{0,1\}^{n \times n}$ is defined as:
\begin{equation}
    W_{ij} = \mathbf{1}\!\bigl(\|\mathbf{s}_i - \mathbf{s}_j\|_2 \leq \delta,\; i \neq j\bigr),
\end{equation}
where $\delta > 0$ is the neighborhood distance threshold. The matrix $\mathbf{W}$ is symmetric with $W_{ii} = 0$. The spatial neighborhood of cell $i$ is denoted by $\partial(i) = \{j : W_{ij} = 1\}$, which serves as the graph structure used to impose the MRF constraint in Step~2.

\subsection{Spatially-Informed Bayesian Hierarchical Model}
\label{sec:step2}

Given $\{\mathbf{A}^{(m)}\}$ and $\mathbf{W}$ from Step~1, we formulate the following Bayesian hierarchical model comprising a Gaussian SBM likelihood, Normal-Gamma conjugate priors, and a spatially-aware partition prior.

\textbf{Generative Model for Cell Similarities.}
We assign a latent state variable $z_i \in \{1, \ldots, k\}$ to represent the spatial domain membership of the $i$-th cell, with the total number of domains $k$ treated as an unknown quantity to be learned from the data. We assume the upper-triangular elements (including the diagonal) of each similarity matrix follow a Gaussian distribution. For any pair of cells $1 \leq i \leq j \leq n$, the likelihood is given by:
\begin{equation}
\label{eq:likelihood}
    A^{(m)}_{ij} \mid z_i, z_j, \mathbf{U}^{(m)}, \mathbf{T}^{(m)}, k
    \;\stackrel{\mathrm{ind}}{\sim}\;
    \mathcal{N}\!\bigl(\mu^{(m)}_{z_iz_j},\;
    (\tau^{(m)}_{z_iz_j})^{-1}\bigr).
\end{equation}
Here, the parameter matrices $\mathbf{U}^{(m)} \in \mathbb{R}^{k \times k}$ and $\mathbf{T}^{(m)} \in \mathbb{R}_+^{k \times k}$ dictate the block-specific expectations ($\mu^{(m)}_{rs}$) and precisions ($\tau^{(m)}_{rs}$) for the $m$-th modality, while maintaining symmetry. Assuming conditional independence of the edges given the domain partition~\citep{holland1983stochastic}, the full likelihood factorizes across the block structure:
\begin{equation}
\label{eq:likelihood-factor}
    P\!\bigl(\mathbf{A}^{(m)} \mid \mathbf{z}, \mathbf{U}^{(m)},
    \mathbf{T}^{(m)}, k\bigr)
    = \prod_{1 \leq r \leq s \leq k}
      P\!\bigl(\mathbf{A}^{(m)}_{[rs]} \mid \mathbf{z}, \mathbf{U}^{(m)},
      \mathbf{T}^{(m)}\bigr),
\end{equation}
where $\mathbf{A}^{(m)}_{[rs]}$ represents the submatrix collecting the similarities between cells situated in domain $r$ and domain $s$.

\textbf{Conjugate Parameter Priors.}
We impose a Normal-Gamma conjugate prior on the mean-precision pairs:
\begin{eqnarray}
\label{eq:prior}
    &&T^{(m)}_{rs} \stackrel{\mathrm{ind}}{\sim} \mathrm{Gamma}(\alpha, \beta),\\
    &&U^{(m)}_{rs} \mid T^{(m)}_{rs}
    \stackrel{\mathrm{ind}}{\sim}
    \mathcal{N}\!\bigl(\mu_0,\; k_0^{-1}(T^{(m)}_{rs})^{-1}\bigr),
    \quad r \leq s.
\end{eqnarray}
We employ a data-driven empirical approach to specify the hyperparameter $\mu_0$~\citep{hu2023bayesian}: for within-domain connections ($r=s$), $\mu_0$ is derived from the average of the diagonal entries of $\mathbf{A}^{(m)}$, whereas for between-domain interactions ($r \neq s$), it takes the average of the off-diagonal entries. The remaining hyperparameters are deterministically set to $k_0 = 10$ and $\alpha = \beta = 1$.

\textbf{Mixture of Finite Mixtures Prior.}
Traditional nonparametric priors, such as the Dirichlet process mixture (conceptualized via the Chinese restaurant process), are known to yield inconsistent estimates of the number of components as the sample size approaches infinity, typically by generating spurious, negligible clusters~\citep{miller2018mixture}. To circumvent this limitation, BaySC adopts an MFM prior to govern the distribution of $k$:
\begin{eqnarray}
\label{eq:mfm}
   && k \sim p(\cdot),
    \quad
    (\pi_1,\ldots,\pi_k)\mid k \sim \mathrm{Dirichlet}(\gamma,\ldots,\gamma),
    \nonumber\\
   && z_i \mid k, \boldsymbol{\pi} \sim \sum_{h=1}^{k} \pi_h\, \delta_h,
    \quad i = 1,\ldots,n.
\end{eqnarray}
We specify $p(\cdot)$ as a zero-truncated Poisson(1) distribution, with $\gamma$ serving as the symmetric Dirichlet concentration parameter. Marginalizing out the mixing weights $\boldsymbol{\pi}$ yields a modified P\'{o}lya urn scheme for the conditional prior of the domain assignments~\citep{miller2018mixture}:
\begin{equation}
\label{eq:polya}
    P(z_i = c \mid \mathbf{z}_{-i})
    \propto
    \begin{cases}
        n_c^{(-i)} + \gamma, & \text{if } c\in\{Ex\},\\[6pt]
        \dfrac{V_n(w+1)}{V_n(w)}\,\gamma, & \text{if } c\in\{New\},
    \end{cases}
\end{equation}
where $n_c^{(-i)}$ signifies the cardinality of domain $c$ excluding the $i$-th cell, $w$ denotes the number of instantiated domains, $\{Ex\} = \{1,\ldots,w\}$ denotes the set of existing domains, and $\{New\} = \{w+1\}$ denotes a new domain. Because the ratio $V_n(w+1)/V_n(w) < 1$, the MFM mechanism inherently penalizes the instantiation of new clusters, guaranteeing an asymptotically consistent estimation of the true domain count.

\textbf{Spatial Regularization via Markov Random Fields.}
Standard MFM formulations implicitly assume exchangeability among cells, failing to account for the spatial autocorrelation fundamental to spatial transcriptomics data. To explicitly capture this geographical dependency, we integrate an MRF~\citep{orbanz2008nonparametric}. Given the spatial neighborhood graph $\mathbf{W}$ from Step~1, we define the energy potential as:
\begin{equation}
\label{eq:energy}
    H(z_i \mid \mathbf{z}_{-i})
    := -\lambda \sum_{j \in \partial(i)} \mathbf{1}(z_j = z_i),
    \quad \lambda \in \mathbb{R}^+,
\end{equation}
imparting an energy reward of $\lambda$ for each spatial neighbor sharing the same domain label as cell $i$.

\textbf{Unified MRFC-MFM Prior Formulation.}
We synthesize the MFM base measure and the MRF structural penalty into a joint prior: $\Pi(\mathbf{z}) \propto P(\mathbf{z})\, M(\mathbf{z})$. Following \citet{hu2023bayesian}, we adapt their theoretical results to our SBM framework, yielding the following label-update probabilities.

\begin{proposition}[Adapted MRFC-MFM Label Update Probabilities, \citep{hu2023bayesian}]
\label{prop:mrfc-mfm-update}
Let $n_c^{(-i)}$ designate the population of domain $c$ excluding cell $i$, and let $K^*$ denote the total number of active domains after the removal of cell $i$. The full conditional probability of assigning cell $i$ to domain $c$ within the Gibbs sampler is:
\begin{equation}
\label{eq:label-full}
    P(z_i = c \mid \mathbf{z}_{-i})
    \propto
    \begin{cases}
        \bigl(n_c^{(-i)} + \gamma\bigr)\,
        \exp\!\Bigl(\lambda \displaystyle\sum_{j \in \partial(i)}
        \mathbf{1}(z_j = c)\Bigr),
        & \text{if } c\in\{Ex\}, \\[10pt]
        \dfrac{V_n(K^*+1)}{V_n(K^*)}\,\gamma,
        & \text{if } c\in\{New\},
    \end{cases}
\end{equation}
where the topological reinforcement term $\exp(\cdot)$ is neutralized when proposing a completely new domain, immunizing novel clusters from the MRF structural penalty.
\end{proposition}

\noindent\textit{The proof of Proposition~\ref{prop:mrfc-mfm-update} is provided in Supplementary Note~2.}

When evaluating an existing domain $c$, the assignment probability combines two forces: a ``rich-get-richer'' dynamic driven by the cluster mass $n_c^{(-i)} + \gamma$, and a topological reinforcement term reflecting the number of spatial neighbors already assigned to $c$. Proposing a new domain confers no spatial energy reward and incurs the MFM deflation penalty $V_n(K^*+1)/V_n(K^*)$. This balance suppresses artifactual isolated domains driven by transcriptomic noise while preserving genuine biological boundaries.

Crucially, the effectiveness of this spatial regularizer hinges on the magnitude of $\lambda$. If $\lambda$ is too small, spatial smoothing is negligible; if excessively large, it forces rigid spatial blocks that override the data likelihood. Drawing on the statistical mechanics of the ferromagnetic Potts model~\citep{wu1982potts}, we establish a heuristic scaling boundary to guide the hyperparameter search space.

\noindent\textbf{Heuristic Scaling of the Spatial Penalty.}
Let $\mathbf{W}$ define a spatial neighborhood graph with average degree $c$, and let $k$ be the current number of active spatial domains. Under a mean-field approximation, the critical threshold scales as $\lambda_c \approx \frac{k}{c}$, delineating two behavioral regimes:
\begin{enumerate}
    \item \textbf{Disordered Regime ($\lambda \ll \lambda_c$):} Domain allocations are primarily governed by the data likelihood and the MFM deflation penalty, potentially leading to fragmented spatial structures.
    \item \textbf{Ordered Regime ($\lambda \ge \lambda_c$):} The MRF induces spontaneous symmetry breaking, heavily biasing the system toward artificially large, homogeneous spatial blocks, overriding the SBM likelihood.
\end{enumerate}

\subsection{Multimodal Fusion and Posterior Inference}
\label{sec:step3}

Conditioned on a specified spatial smoothing penalty $\lambda$ and neighborhood distance threshold $\delta$, Step~3 executes the Bayesian hierarchical model formulated in Step~2. This inference stage comprises four components: multimodal weighted log-likelihood fusion, a new-domain proposal mechanism, a collapsed Gibbs sampling algorithm, and posterior point estimation with cell-level uncertainty quantification.

\textbf{Multimodal Weighted Log-Likelihood.}
All modalities share a unified domain label vector $\mathbf{z}$ while maintaining modality-specific block parameter sets $\{\mathbf{U}^{(m)}, \mathbf{T}^{(m)}\}$. Building on Proposition~\ref{prop:mrfc-mfm-update}, the log-probability of assigning cell $i$ to an existing domain $c$ is:
\begin{eqnarray}
\label{eq:fusion}
    \log P(z_i = c \mid \cdot)
    &=& \sum_{m} \alpha_m \cdot \ell^{(m)}(i, c)\nonumber\\
    &&+ \lambda \sum_{j \in \partial(i)} \mathbf{1}(z_j = c)
    + \log\!\bigl(n_c^{(-i)} + \gamma\bigr),
\end{eqnarray}
balancing three terms: the weighted multimodal data likelihood, the MRF spatial coherence reward, and the MFM ``rich-get-richer'' effect. The conditional log-likelihood of modality $m$ for domain $c$ is:
\begin{equation}
\label{eq:modality-loglik}
    \ell^{(m)}(i, c)
    = \sum_{j \neq i}
      \Bigl[
        \tfrac{1}{2}\log\tau^{(m)}_{c,z_j}
        - \tfrac{\tau^{(m)}_{c,z_j}}{2}
          \bigl(A^{(m)}_{ij} - \mu^{(m)}_{c,z_j}\bigr)^2
      \Bigr],
\end{equation}
where when $z_j = c$ the within-domain parameters $(\mu^{(m)}_{cc},\,\tau^{(m)}_{cc})$ apply, and when $z_j = s \neq c$ the cross-domain parameters $(\mu^{(m)}_{cs},\,\tau^{(m)}_{cs})$ apply. The weights $\alpha_m \geq 0$ govern the relative contribution of each modality. For single-modality data, the framework reduces to $M=1$ and $\alpha_1 = 1$.

\textbf{New-Domain Proposal Mechanism.}
\label{sec:newcluster}
When cell $i$ is proposed to instantiate a novel domain $c = K^*+1$, no existing members are available, rendering the standard block likelihoods undefined. In this case, the algorithm defaults to the Normal-Gamma marginal likelihood, using the cell's self-similarity diagonal entry $A^{(m)}_{ii}$ as the sole observation:
\begin{eqnarray}
\label{eq:new-domain}
    \log P(z_i = K^*+1 \mid \cdot)
    &=& \sum_{m} \alpha_m \cdot \log m^{(m)}\!\bigl(A^{(m)}_{ii}\bigr)
    + \log\gamma\nonumber\\
    && + \log V_n(K^*+1) - \log V_n(K^*),
\end{eqnarray}
where the analytical marginal likelihood is:
\begin{eqnarray}
\label{eq:marginal-lik}
    \log m^{(m)}(A^{(m)}_{ii})
    &=& \log\Gamma(\alpha_n) - \log\Gamma(\alpha)\nonumber\\
    &&+ \alpha\log\beta - \alpha_n\log\beta_n
    + \tfrac{1}{2}\log\tfrac{k_0}{k_n}\,,
\end{eqnarray}
with updated hyperparameters $k_n = k_0 + 1$, $\alpha_n = \alpha + \tfrac{1}{2}$, and $\beta_n = \beta + \dfrac{k_0}{2k_n}\bigl(A^{(m)}_{ii} - \mu_0^{\mathrm{diag}}\bigr)^2$. Because diagonal entries $A^{(m)}_{ii}$ deviate substantially from the off-diagonal range after Fisher-Z transformation, this marginal likelihood is inherently low, providing a data-driven penalty against unnecessary domain creation that compounds the MFM deflation term $V_n(K^*+1)/V_n(K^*)$.

\textbf{Collapsed Gibbs Sampling.}
Exploiting Normal-Gamma conjugacy, all continuous block parameters $\{\mathbf{U}^{(m)}, \mathbf{T}^{(m)}\}$ are analytically marginalized out, restricting the MCMC sampling space to the discrete labels $\mathbf{z}$ and circumventing the mixing issues associated with reversible-jump MCMC in trans-dimensional spaces~\citep{hu2023bayesian}. Each iteration alternates between two updates:

First, for each cell $i$, a new label is sampled from $K^*+1$ candidates via Eq.~\eqref{eq:fusion} for existing domains and Eq.~\eqref{eq:new-domain} for a new domain. Empty domains are immediately purged and labels renumbered.

Second, block parameters are resampled. Let $\mathcal{D}^{(m)}_{rs} = \{A^{(m)}_{ij} : z_i=r,\, z_j=s,\, i \leq j\}$ with cardinality $n_{rs}$, sample mean $\bar{A}^{(m)}_{rs}$, and residual sum of squares $\mathrm{SSE}^{(m)}_{rs} = \sum_{A \in \mathcal{D}^{(m)}_{rs}}(A - \bar{A}^{(m)}_{rs})^2$. The posterior hyperparameters update as:
\begin{align}
\label{eq:posterior-params}
    k_n &= k_0 + n_{rs}, \quad
    \mu_n = \frac{k_0\,\mu_0 + n_{rs}\,\bar{A}^{(m)}_{rs}}{k_n}, \quad
    \alpha_n = \alpha + \frac{n_{rs}}{2}\,, \nonumber\\
    \beta_n &= \beta + \frac{1}{2}\!\left(
      \mathrm{SSE}^{(m)}_{rs}
      + \frac{n_{rs}\,k_0}{k_n}
      \bigl(\bar{A}^{(m)}_{rs} - \mu_0\bigr)^2
    \right),
\end{align}
and fresh block parameters are drawn via:
\begin{equation}
\label{eq:posterior-sample}
    \tau^{(m)}_{rs} \sim \mathrm{Gamma}(\alpha_n, \beta_n),\quad
    \mu^{(m)}_{rs} \mid \tau^{(m)}_{rs}
    \sim \mathcal{N}\!\bigl(\mu_n,\; (k_n\,\tau^{(m)}_{rs})^{-1}\bigr).
\end{equation}

\textbf{Posterior Point Estimation and Uncertainty Quantification.}
Following burn-in, $M$ posterior samples are collected. To resolve label switching, we apply the approach of \citet{dahl2006model}. At iteration $t$, define the binary co-membership matrix $\mathbf{B}^{(t)}$ with entries $B^{(t)}_{ij} = \mathbf{1}(z^{(t)}_i = z^{(t)}_j)$, and compute the posterior mean adjacency matrix $\bar{B} = \frac{1}{M}\sum_{t=1}^{M} \mathbf{B}^{(t)}$. The representative sample is selected as:
\begin{equation}
\label{eq:dahl}
    t^* = \arg\min_{t \in \{1,\ldots,M\}}
    \sum_{i,j}
    \bigl(B^{(t)}_{ij} - \bar{B}_{ij}\bigr)^2.
\end{equation}
The optimal domain number $K$ and all parameter estimates are extracted from the $t^*$-th sample. The matrix $\bar{B}$ also provides cell-level uncertainty: $\bar{B}_{ij} \approx 1$ indicates near-certain co-assignment, $\bar{B}_{ij} \approx 0$ indicates clear separation, and intermediate values mark ambiguous boundaries. To quantify uncertainty at the single-cell level, let $\mathcal{C}_c = \{j : \hat{z}_j = c\}$ be the set of cells assigned to domain $c$ under the point estimate $\hat{\mathbf{z}}$. The mean posterior co-membership affinity of cell $i$ to its assigned domain is:
\begin{equation}
\label{eq:affinity}
    \bar{p}_{i,c}
    = \frac{1}{|\mathcal{C}_c \setminus \{i\}|}
      \sum_{j \in \mathcal{C}_c,\, j \neq i}
      \bar{B}_{ij},
\end{equation}
and the cell-level uncertainty score is:
\begin{equation}
\label{eq:uncertainty}
    u_i = 1 - \max_{c}\,\bar{p}_{i,c} \;\in\; [0,1],
\end{equation}
where $u_i \approx 0$ denotes confident domain membership and $u_i \approx 1$ identifies putative boundary cells, transitional progenitors, or noisy captures.

\subsection{Data-Driven Hyperparameter Selection via mDIC}
\label{sec:step4}

The performance of Step~3 relies on two hyperparameters: the neighborhood distance threshold $\delta$, which defines the physical radius of cell--cell interactions and determines the density of $\mathbf{W}$, and the spatial penalty $\lambda$, which controls the rigidity of domain boundaries. Since tissue architectures vary across biological samples, from stratified cortical layers to disorganized tumor microenvironments, default parameters are sub-optimal. Step~4 therefore employs the mDIC to jointly select the optimal pair $(\lambda^*, \delta^*)$ via a data-driven grid search.

\textbf{Theoretical Search Space and Candidate Grid.}
We construct the candidate grid $\mathcal{G}$ using the theoretical bounds from Step 2. Because the Markov chain risks collapsing into an artifactually frozen state when $\lambda \ge \lambda_c$, we restrict $\lambda$ to $[0, \lambda_{\max}]$, where $\lambda_{\max} < \lambda_c = k/c$. For each configuration $(\lambda, \delta) \in \mathcal{G}$, the complete MCMC inference pipeline is executed independently, and the optimal parameters are selected by:
\begin{equation}
\label{eq:grid-select-mdic}
    (\lambda^*, \delta^*) =
    \arg\min_{(\lambda, \delta) \in \mathcal{G}} \mathrm{mDIC}(\lambda, \delta).
\end{equation}

\textbf{The mDIC Formulation.}
The standard DIC~\citep{spiegelhalter2002bayesian} calibrates its penalty on an AIC-like scale, which under-penalizes complexity in large-scale spatial transcriptomics data, leading to the proliferation of biologically meaningless micro-domains. We instead employ the mDIC, which adjusts the penalty to a BIC-consistent scale~\citep{schwarz1978estimating}:
\begin{equation}
\label{eq:mdic}
    \mathrm{mDIC}
    = \overline{D(\boldsymbol{\theta})}
    + \log\!\left(\frac{n(n+1)}{2}\right) p_D\,,
\end{equation}
where the posterior mean deviance is approximated from the $M$ post-burn-in Gibbs samples:
\begin{equation}
\label{eq:mean-deviance}
    \overline{D(\boldsymbol{\theta})}
    = -2\,\mathbb{E}_{\mathrm{post}}\!\Bigl[
        \log P\!\bigl(\{\mathbf{A}^{(m)}\} \mid \boldsymbol{\theta}\bigr)
      \Bigr],
\end{equation}
and the effective complexity is:
\begin{equation}
\label{eq:effective-params}
    p_D = \overline{D(\boldsymbol{\theta})} - D(\hat{\boldsymbol{\theta}}),
\end{equation}
with $D(\hat{\boldsymbol{\theta}})$ the deviance at the Dahl posterior point estimate (Eq.~\eqref{eq:dahl}). In practice, $p_D$ measures posterior dispersion: unstable MCMC chains yield inflated $p_D$ values. The penalty multiplier $\log\!\bigl(\frac{n(n+1)}{2}\bigr)$ reflects the effective sample size of the Gaussian SBM---the number of upper-triangular elements in $\mathbf{A}^{(m)}$---scaling the penalty conservatively with the $O(n^2)$ observation space to guard against overfitting.

\textbf{The Non-Spatial Safeguard.}
The candidate grid includes a $\lambda = 0$ baseline. If a tissue sample lacks genuine spatial autocorrelation, the mDIC will select $\lambda^* = 0$, reducing the model to a standard unstructured MFM. This ensures spatial contiguity is enforced only when empirically supported by the data, avoiding the over-smoothing artifacts common in GNN-based methods.

\section{Results}

\subsection{Experimental Setup}

\textbf{Datasets.}
We evaluated BaySC on twelve spatial transcriptomics datasets spanning both single-modality and multi-modality settings; detailed dataset statistics are provided in Supplementary Note~3. Single-modality experiments used three benchmark sources: \textbf{Human Breast Cancer (HBC)}, a 10x~Visium dataset of human breast cancer tissue annotated with 20 spatial domains based on H\&E morphology and pathological features ($n=3{,}798$ spots); \textbf{STARmap}, a spatial transcriptomics dataset of the mouse visual cortex comprising 7 spatial domains ($n=1{,}207$ cells); and \textbf{HER2-positive Breast Cancer}, comprising eight tissue sections (A1--H1) with 3--6 annotated domains and 167--659 spots per section. Multi-modality experiments used two paired spatial multi-omics datasets. \textbf{Human Lymph Node A1}, a 10x~Visium dataset jointly profiling RNA and protein (ADT) modalities ($n=3{,}484$ spots). The original H\&E-based manual annotation defines 10 anatomical compartments, further merged into 6 broader classes for clustering evaluation, following the comparison setting adopted in prior analyses of this dataset. \textbf{MISAR-seq Mouse E15.5 Brain}, jointly profiling spatial mRNA and ATAC-seq ($n=1{,}949$ spots, $K=7$).

\textbf{Data Preprocessing.}
Raw count matrices for each modality were preprocessed through standard pipelines to obtain low-dimensional embeddings. RNA profiles were normalized, log-transformed, and reduced to 50 principal components via PCA; ATAC profiles were binarized, TF-IDF transformed, and reduced to 50 components via LSI with the first component discarded; protein/ADT profiles were CLR-normalized and reduced to 30 principal components via PCA. All embeddings were subsequently standardized per cell ($z$-score, $\mathrm{ddof}=1$) before input into BaySC.

\textbf{Evaluation Metrics.}
We evaluate spatial clustering performance using six complementary metrics, whose formal definitions are given in Supplementary Note 4. Specifically, we report four standard clustering metrics (ARI, AMI, NMI, and Homogeneity), as well as two spatially aware metrics (Moran's I and spARI~\citep{yan2025spari}) to quantify spatial coherence. Notably, spARI assigns asymmetric penalties to misclassified cell pairs according to their Euclidean distances in tissue space, making it particularly well aligned with the biological assumption of spatial continuity in tissue domains. We therefore regard spARI as the primary metric.

\begin{figure*}[!t]
\centering
\includegraphics[width=0.9\linewidth]{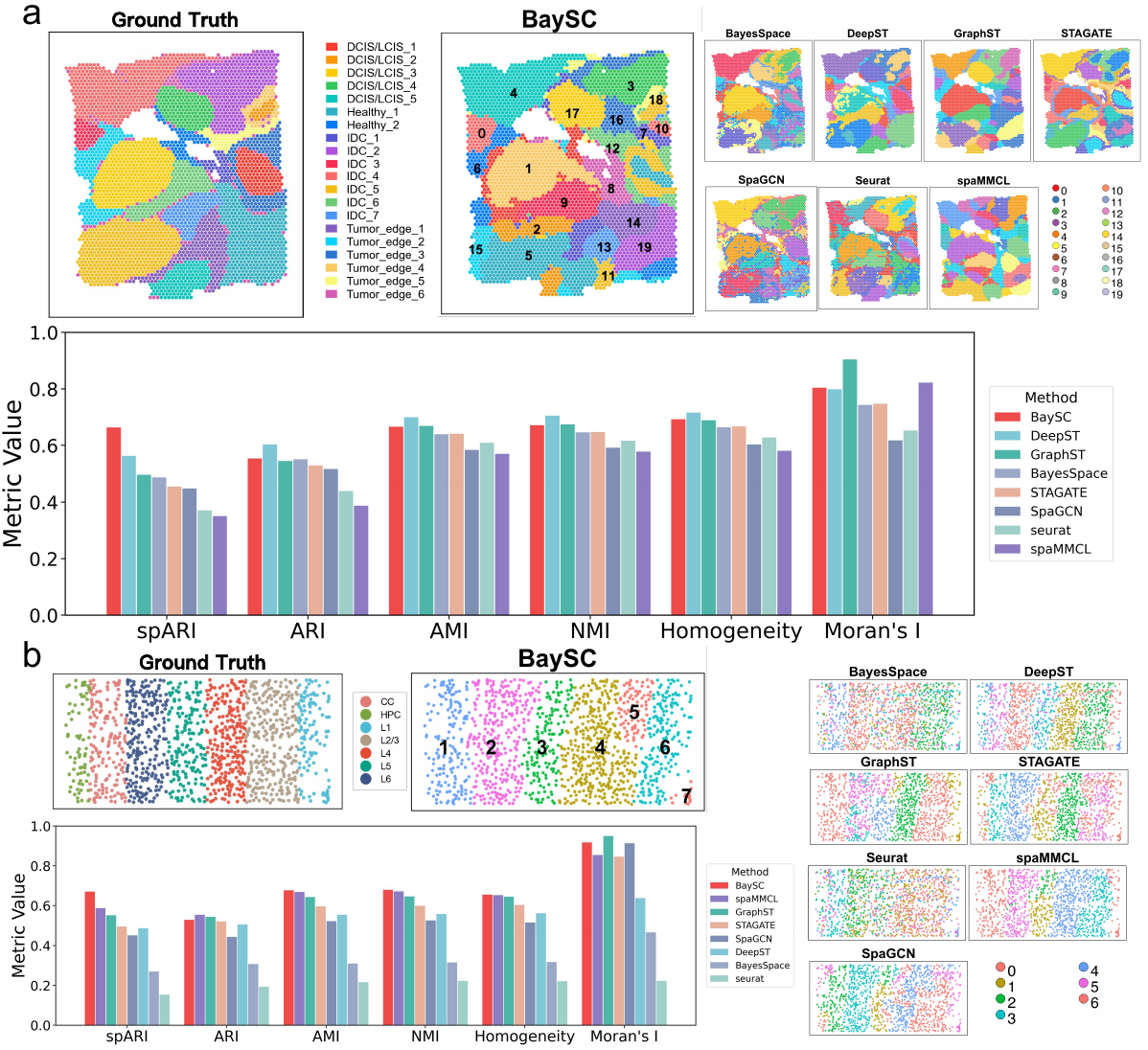}
\caption{Single-modality spatial clustering results on the Human Breast Cancer 10x~Visium dataset and the STARmap Mouse Visual Cortex dataset.
\textbf{(a)}~Human Breast Cancer ($n=3{,}798$ spots, $K=20$ domains). From left to right: ground-truth domain annotations comprising 20 regions (DCIS/LCIS, IDC, tumour edge, and healthy tissue); predicted spatial domains by BaySC; and spatial domain maps produced by seven competing methods (BayesSpace, DeepST, GraphST, STAGATE, SpaGCN, Seurat, spaMMCL). Bar charts below compare ARI, AMI, NMI, Homogeneity, and Moran's~I across all methods, with BaySC shown in red.
\textbf{(b)}~STARmap Mouse Visual Cortex ($n=1{,}207$ cells, $K=7$ domains). From left to right: ground-truth cortical layer annotations (CC, HPC, L1--L6); predicted spatial domains by BaySC; and spatial domain maps produced by seven competing methods. Bar charts below compare all six evaluation metrics across methods, with BaySC shown in red.}
\label{fig:cancer_starmap}
\end{figure*}

\subsection{Single-Modality Spatial Domain Identification}

\textbf{Human Breast Cancer.}
We evaluated BaySC on the HBC dataset ($n=3798$ spots, $K=20$). As shown in Fig.~\ref{fig:cancer_starmap}a, BaySC ranked first in spARI with a score of 0.6643, exceeding the second-best method by 0.101. BaySC also attained ARI, AMI, NMI, and Homogeneity scores of 0.5541, 0.6663, 0.6722, and 0.6928, respectively, all of which are comparable to those of the baseline methods. The superior spARI performance indicates that the label-layer MRF constraint in BaySC effectively preserves the spatial continuity of predicted domains within the tumor microenvironment, while leaving the probabilistic modeling of expression features largely unaffected. Notably, BaySC automatically inferred the number of domains as $K=20$ without any manual specification.

\textbf{STARmap Mouse Visual Cortex.}
On the STARmap dataset ($n=1{,}207$ cells, $K=7$), BaySC ranked first in AMI (0.6772), NMI (0.6799), and Homogeneity (0.6555), second in ARI (0.5291; spaMMCL achieved 0.5545), and first in spARI with a score of 0.6708, outperforming the second-best method, spaMMCL (0.5879). BaySC also achieved the highest Moran's~I value among all methods (0.9182; Fig.~\ref{fig:cancer_starmap}b). The mouse visual cortex exhibits a clear laminar organization (CC, HPC, and L1--L6), characterized by strong spatial continuity across layers and distinct transcriptomic profiles between them. The high Moran's~I value indicates that the domain boundaries predicted by BaySC are highly spatially coherent, while the strong AMI and NMI scores show that the transcriptomic specificity of individual cortical layers is also well distinguished. This balanced performance directly reflects the design of BaySC, in which the label-layer MRF constraint and the SBM likelihood are specified independently.

\begin{figure*}[!t]
\centering
\includegraphics[width=0.9\linewidth]{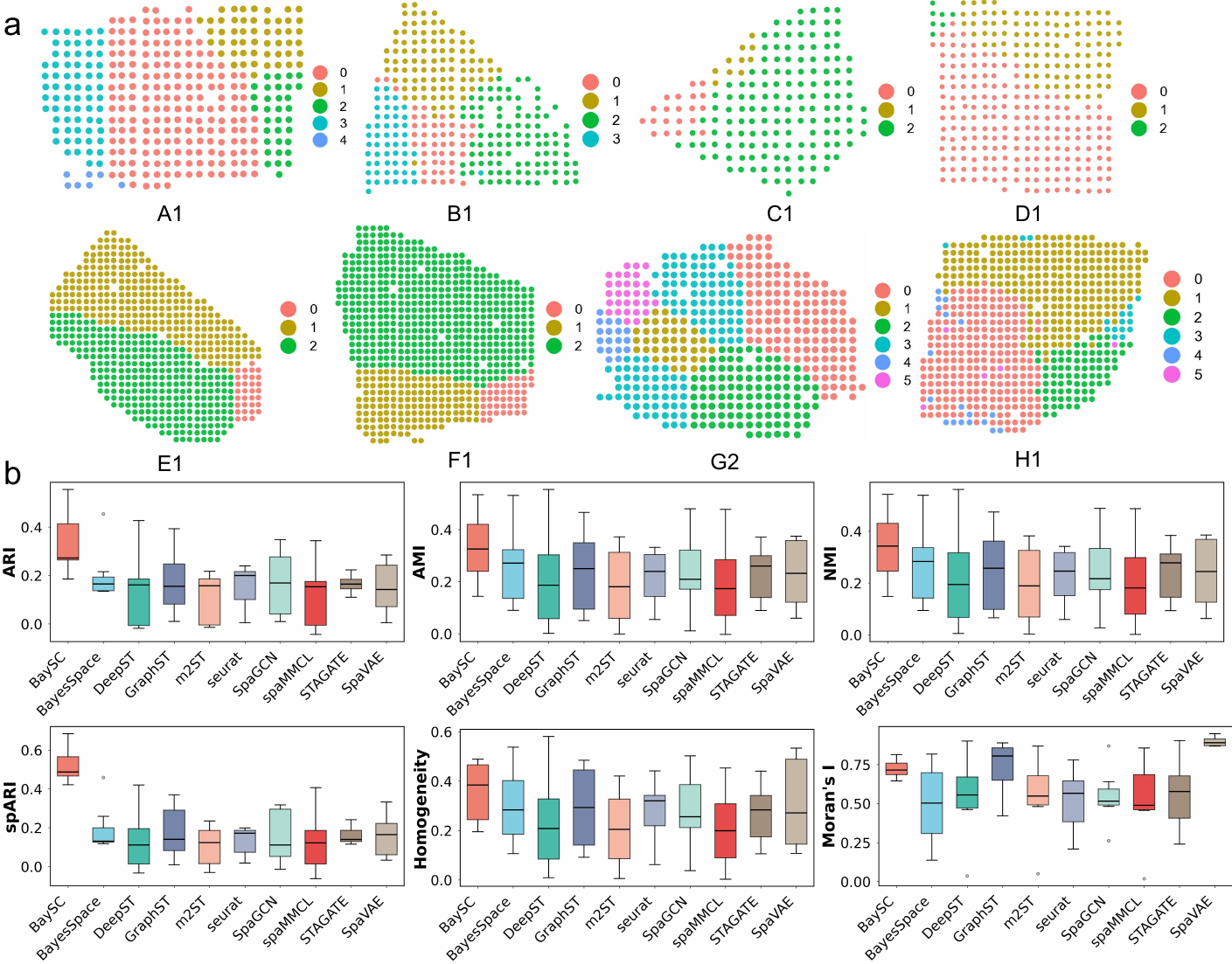}
\caption{Spatial clustering results on the HER2-positive Breast Cancer dataset across eight tissue sections (A1--H1).
\textbf{(a)}~Predicted spatial domains by BaySC for each of the eight sections, with annotated domain counts $K=3$--$6$ and spot counts $n=167$--$659$.
\textbf{(b)}~Box plot comparison of ARI, AMI, NMI, spARI, Homogeneity, and Moran's~I across all eight sections, with BaySC shown in the first position. Per-section numerical results and clustering maps are provided in Supplementary Note~5.}
\label{fig:her2}
\end{figure*}

\textbf{HER2-positive Breast Cancer.}
To examine the robustness of BaySC across tissue slices of varying complexity, we evaluated it on eight HER2-positive breast cancer sections (A1--H1, $K=3$--$6$, $n=167$--$659$). As shown in Fig.~\ref{fig:her2}b, BaySC achieved the highest median spARI with the smallest cross-slice variation. Its median ARI, AMI, and NMI scores also ranked among the best across the methods. The advantage was particularly significant on more challenging slices with weaker signals. On slice E1, BaySC achieved an spARI of 0.4674, whereas all baseline methods scored below 0.15. On slice F1, BaySC reached 0.4215, while the second-best method, STAGATE, achieved only 0.1394. On slice A1, BaySC reached 0.4871, compared with only 0.1956 for the second-best method, Seurat. These results suggest that the systematic penalty imposed by the MFM prior on the introduction of new domains enables BaySC to maintain robust spatial domain identification even under sparse data and weak signal conditions. Detailed results are presented in Supplementary Note~5.

\subsection{Multi-Modal Spatial Domain Identification}

\begin{figure*}[!t]
\centering
\includegraphics[width=0.9\linewidth]{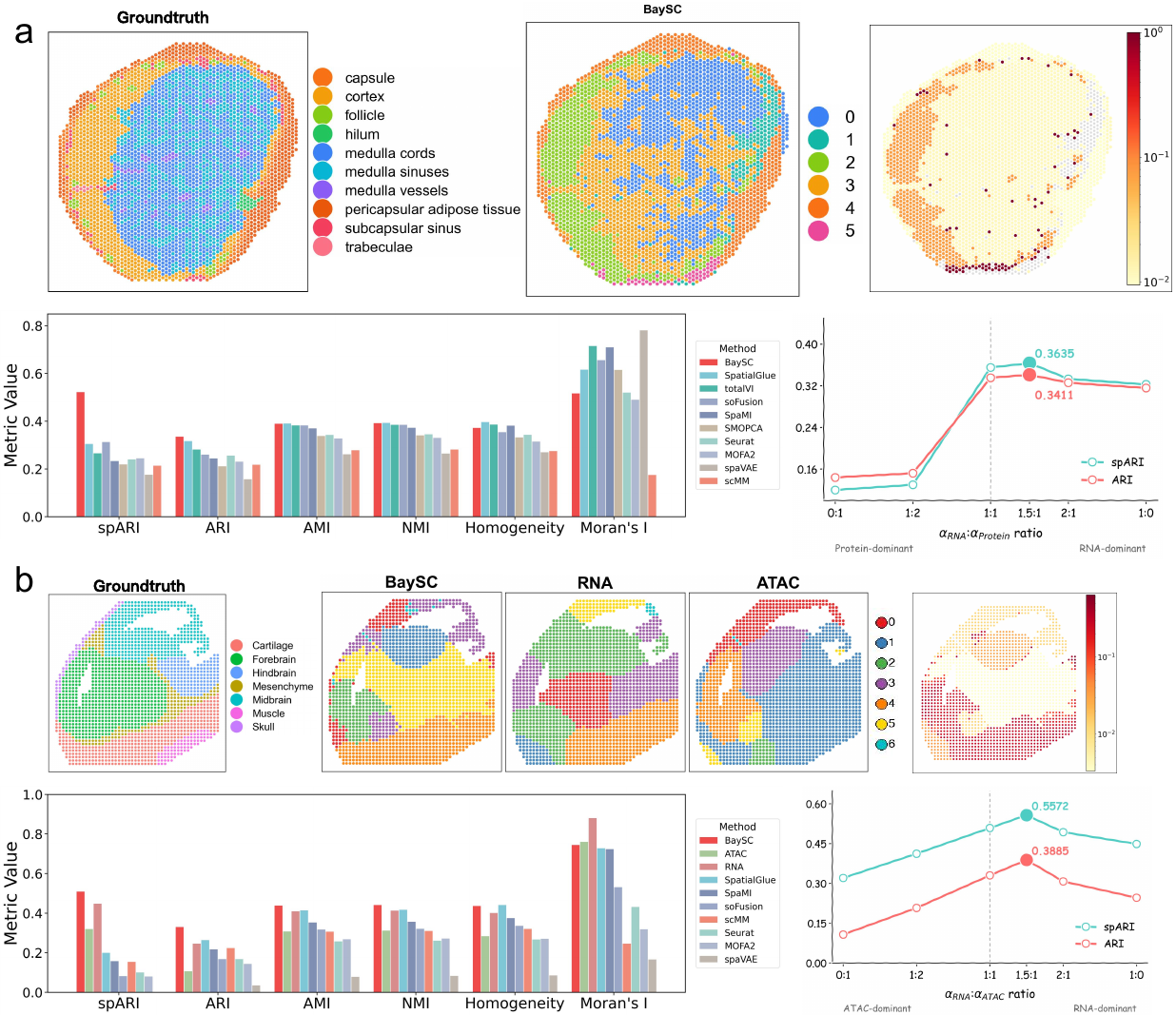}
\caption{Multi-modal spatial clustering results on the Human Lymph Node A1 and MISAR-seq Mouse E15.5 Brain datasets.
\textbf{(a)}~Human Lymph Node A1 ($n=3{,}484$ spots, $K=6$ domains, RNA\,+\,Protein modalities). From left to right: ground-truth annotations (ten structural compartments merged into six classes); predicted spatial domains by BaySC; posterior uncertainty map ($\log_{10}$ scale); bar charts comparing six metrics across ten methods; and sensitivity analysis of the $\alpha_{\text{RNA}}{:}\alpha_{\text{Protein}}$ weight ratio (peak: spARI\,$=$\,0.3635, ARI\,$=$\,0.3411).
\textbf{(b)}~MISAR-seq Mouse E15.5 Brain ($n=1{,}949$ spots, $K=7$ domains, RNA\,+\,ATAC modalities). From left to right: ground-truth annotations; predicted domains by BaySC (joint RNA\,+\,ATAC), RNA-only baseline, and ATAC-only baseline; posterior uncertainty map; bar charts comparing all six metrics; and sensitivity analysis of the $\alpha_{\text{RNA}}{:}\alpha_{\text{ATAC}}$ weight ratio (peak near 1.5:1; spARI\,$=$\,0.5572, ARI\,$=$\,0.3885).}
\label{fig:multimodal}
\end{figure*}

\textbf{Human Lymph Node A1.}
We evaluated BaySC on the Human Lymph Node A1 dataset ($n=3484$ spots, $K=6$; bimodal RNA + protein ADT data). As shown in Fig.~\ref{fig:multimodal}a, BaySC achieved the highest ARI (0.3354), while its AMI (0.3902) and NMI (0.3927) were comparable to those of the strongest baseline methods. BaySC further attained an spARI of 0.5221, exceeding the second-best method by 0.209, which is the largest margin observed between BaySC and the runner-up across all datasets considered in this study. The resulting clustering map accurately delineates the major structural compartments of the lymph node and shows high agreement with the reference annotation.

The posterior uncertainty map visualizes spot-wise assignment uncertainty using the $\log_{10}(u_i)$ color scale. Spots located in the subcapsular sinus and trabeculae exhibit markedly elevated uncertainty scores $u_i$ (Eq.~\eqref{eq:uncertainty}), consistent with the intrinsically ambiguous transition boundaries of these structures in histology. By contrast, spots in regions with well-defined transcriptomic signatures, such as the cortex and follicle, show near-zero uncertainty and correspondingly high assignment confidence. This spot-level uncertainty pattern is biologically coherent and offers a practical basis for downstream analysis: high-confidence spots may be prioritized for mechanistic investigation, while additional biological validation can be focused on high-uncertainty boundary regions.

We further conducted a sensitivity analysis on the ratio $\alpha_{\text{RNA}}{:}\alpha_{\text{Protein}}$ (right panel of Fig.~\ref{fig:multimodal}a). Both ARI and spARI peaked when the RNA modality was assigned a slightly higher weight than the protein modality (spARI = 0.3635, ARI = 0.3411), and both metrics declined as the weight ratio became increasingly skewed. The performance of either single modality alone was substantially inferior to that of bimodal integration, indicating that both modalities are indispensable for resolving lymph node tissue architecture.

\textbf{MISAR-seq Mouse E15.5 Brain.}
We evaluated BaySC on the MISAR-seq Mouse E15.5 Brain dataset ($n=1{,}949$ spots, $K=7$; bimodal RNA + ATAC data). As shown in Fig.~\ref{fig:multimodal}b, BaySC ranked first in ARI (0.3305), AMI (0.4391), NMI (0.4423), Moran's~I (0.7455), and spARI (0.5091). Notably, the spARI values of all other multimodal methods remained below 0.23 (SpatialGlue: 0.1998; SMOPCA: 0.2258), placing them more than 0.25 behind BaySC.

Comparison with unimodal baselines further highlights the necessity of multimodal integration. The RNA-only baseline (ARI 0.2462, spARI 0.4488) recovered the overall outlines of the major brain regions, but several boundaries remained indistinct. In contrast, the ATAC-only baseline (ARI 0.1074, spARI 0.3206) produced substantial misclassification in the Cartilage and Skull regions, indicating that chromatin accessibility information alone is insufficient to distinguish these structures. By jointly integrating both modalities, BaySC outperformed both unimodal baselines in terms of ARI and spARI, and accurately resolved all seven anatomical regions (Fig.~\ref{fig:multimodal}b).

The posterior uncertainty map (Fig.~\ref{fig:multimodal}b) further shows that spots near the boundaries between the forebrain, midbrain, and hindbrain exhibit markedly elevated $u_i$ values. This pattern is highly consistent with the developmental transition gradients characteristic of these regions at embryonic day E15.5: progenitor cells near fate-decision boundaries have not yet fully differentiated at the transcriptomic level and therefore exhibit inherent ambiguity in clustering. These results suggest that the uncertainty quantification provided by BaySC can capture cell subpopulations with genuine developmental significance, thereby offering a computational basis for prioritizing targeted experimental validation.

Sensitivity analysis of the ratio $\alpha_{\text{RNA}}{:}\alpha_{\text{ATAC}}$ showed a similar pattern: performance peaked when the RNA modality was assigned a slightly larger weight (1.5:1; spARI = 0.5572, ARI = 0.3885), and declined when either modality was overly emphasized. Taken together with the results from both multimodal datasets, these findings indicate that RNA plays a relatively dominant role in spatial domain identification, while the complementary information provided by the other modality remains essential. Unlike deep learning methods where modality weights are typically learned implicitly by the network, the explicit parameter $\alpha_m$ offers a transparent and quantifiable tool for analyzing modality contributions and for conducting evidence-based weight tuning across different tissue contexts.

\section{Discussion and Future Work}

In this study, we introduced BaySC, a fully Bayesian integrative framework for spatial domain identification in spatial transcriptomics and multi-omics data. By combining an MFM prior with an MRF constraint applied to discrete latent label assignments, BaySC decouples data-driven inference of the cluster number from the enforcement of spatial contiguity. This design prevents the feature-blurring artifacts of GNN-based approaches, enabling precise delineation of biological boundaries without distorting the underlying expression geometry. Coupled with data-driven hyperparameter calibration via mDIC and a weighted log-likelihood formulation for multimodal fusion, BaySC provides a principled, end-to-end generative model. Benchmarking confirmed competitive performance on standard clustering metrics and consistent superiority on the spatially aware spARI metric.

Beyond its algorithmic contributions, BaySC offers direct utility for biological interpretation through uncertainty quantification. Unlike deterministic deep learning methods that produce rigid boundaries and opaque representations, BaySC leverages the Dahl method to extract a posterior co-membership matrix, allowing researchers to explicitly quantify clustering confidence at the single-cell or spot level. This enables systematic identification of ambiguous spatial boundaries, including developmental gradients in embryogenesis and invasive margins in tumor microenvironments, that conventional algorithms force into artificial discrete categories. The modality-specific weights further inform researchers about which molecular layers, such as epigenomic accessibility versus protein abundance, drive local tissue organization, yielding biological insights into spatially heterogeneous multi-omics data.

Despite its theoretical rigor and empirical success, BaySC has several methodological limitations. First, while analytical marginalization streamlines the collapsed Gibbs sampler, MCMC inference on atlas-scale datasets remains computationally intensive relative to optimized neural network forward passes. Second, the static, distance-based neighborhood graph defined by $\delta$ may under-connect sparse regions or over-connect dense structures in tissues with highly variable cellular densities. Finally, the Fisher-Z transformation implicitly assumes approximately linear pairwise relationships in the latent space, and may not fully capture highly non-linear transcriptomic manifolds without additional pre-conditioning.

These limitations suggest clear directions for future work. To address computational bottlenecks at whole-organ scale, we plan to explore variational inference approximations or split-merge MCMC strategies that retain the theoretical guarantees of the MFM prior while accelerating convergence. As spatial technologies advance from 2D sections to 3D volumes and spatio-temporal tracking, the MRF component can be naturally extended to multi-slice or temporal spatial graphs, enabling 3D/4D spatial domain identification. Finally, adaptive graph construction with edge weights learned from local tissue morphology or cellular density could further enhance BaySC's resolution of single-cell spatial omics data.

\section*{Conflicts of Interest}
The authors declare that they have no competing interests.

\section*{Funding}
This work is supported in part by funds from the National Science Foundation (NSF: \# 1636933 and \# 1920920) and RGF grant (No. 14300425) from the Research Grant Council of Hong Kong.

\section*{Data Availability}
All spatial transcriptomics and spatial multi-omics datasets used in this study are publicly available, as summarized in Supplementary Note~3.

\section*{Author Contributions}
H.N. and G.H. conceived the initial idea, supervised this study, and acquired the funding. X.L. developed the methodology, designed the model framework, implemented the software, performed all experiments, and wrote the original draft. X.D. reproduced the baseline methods and assisted with the experiments. Z.D. designed and produced the visualizations. L.S. and X.W. contributed to the methodological design. L.S., X.W., X.S., H.N., and G.H. reviewed and revised the manuscript. All authors have read and approved the final manuscript.

\section*{Acknowledgments}
The authors thank the anonymous reviewers for their valuable suggestions. AI-assisted language editing was performed using Claude (Anthropic).

\bibliographystyle{unsrtnat}
\bibliography{reference}

\end{document}